\documentclass[pdflatex,sn-aps]{sn-jnl}

\usepackage{comment}
\usepackage{graphicx}%
\usepackage{multirow}%
\usepackage{amsmath,amssymb,amsfonts}%
\usepackage{amsthm}%
\usepackage{mathrsfs}%
\usepackage[title]{appendix}%
\usepackage{xcolor}%
\usepackage{textcomp}%
\renewcommand\d{\mathrm{d}}
\newcommand\e{\mathrm{e}}
\newcommand\vr{\boldsymbol{r}}
\newcommand\der[3][]{\frac{\d^{#1} #2}{\d{#3}^{#1}}}
\newcommand\avg[1]{\left\langle{#1}\right\rangle}

\theoremstyle{thmstyleone}%
\theoremstyle{thmstyletwo}%

\theoremstyle{thmstylethree}%

\begin{document}

\title[Article Title]{Statistical properties of the gravitational force through ordering statistics}


\author*[1]{\fnm{Constantin} \sur{Payerne}}\email{constantin.payerne@gmail.com}

\author[2]{\fnm{Vincent} \sur{Rossetto}}

\affil[1]{Université Paris-Saclay, CEA, IRFU, 91191, Gif-sur-Yvette, France}

\affil[2]{Université Grenoble Alpes, CNRS, LPMMC, 38000, Grenoble, France}

\abstract{We study the statistical distribution of Newtonian gravitational forces acting on a test particle embedded in an infinite, homogeneous, and uncorrelated random gas of particles. In the limit where both the number of neighboring particles and the confining volume tend to infinity with constant density, this distribution converges to the classic Holtsmark distribution. Our focus here is on the contribution of the nearest particle neighbors to the total Newtonian force. To this end, we derive the joint spatial distribution of the nearest neighbors in arbitrary spatial dimensions, and show that, in three dimensions, the divergence of the Holtsmark distribution’s variance originates entirely from the dominant influence of the nearest neighbor.}


\maketitle

\section{Introduction}
\label{sec:introduction}
In this paper, we investigate the statistical properties of the gravitational force generated by a random distribution of field sources (such as stars or galaxies). A central result in this context is the Holtsmark distribution \cite{holtsmark19}, which characterizes the probability distribution of the gravitational force acting on a test particle within an infinite, homogeneous, and uncorrelated particle ensemble. This distribution offers a foundational description of stochastic long-range interacting fields, with important applications in astrophysics \cite{chandra42, chandra43, Pietronero_2002, Sylos_Labini_2008} and plasma physics \cite{salzmann, demura10, gigosos14, dyachkov98}. In three spatial dimensions \cite{chavanis09}, the Holtsmark distribution exhibits a formally divergent variance, arising from the dominant influence of the nearest neighbor.
Our goal in this work is to revisit the statistical nature of the gravitational force using tools from order statistics, particularly the distribution of the $n$-th nearest neighbor. This approach enables us to clarify the origin of the divergence in the Holtsmark distribution and to disentangle the relative roles of local versus distant contributions to the total force.

The paper is organized as follows; In Section \ref{sec:stat_distances}, we derive the distribution of the $n$-th nearest neighbor in a homogeneous random medium. The first main result of this paper is the $n$-th nearest neighbor radial distribution in Eq. \eqref{eq:wn}, the second main result is the joint spatial distribution of the $n<N$ nearest neighbors in Eq. \eqref{eq:wN}. Section \ref{sec:gravity_homogeneous_media} then applies these results to analyze the gravitational contribution of the $n$-th nearest neighbor, with particular emphasis on its connection to the properties of the Holtsmark distribution. We conclude in Section \ref{sec:conclusion}.

\section{Statistics of distances between particles}
\label{sec:stat_distances}
\subsection{Nearest neighbors}
We begin by considering a cloud of particles with uniform spatial distribution and mean density~$\rho$. The key statistical property of such a system follows from the fact that the expected number of particles within a volume~$V$ is $\rho V$. Consequently, the actual number of particles contained in~$V$ is a random variable distributed according to a Poisson law \cite{Poisson1837} with mean~$\rho V$:
$$\mathbb{P}\{ \mathcal N(V)=n \} = \frac{(\rho V)^n}{n!}\mathrm{e}^{-\rho V}
$$
where $\mathcal N(V)$ is the random variable counting the particles in a volume~$V$. More calculation details and properties of the Poisson distribution are shown in Appendix \ref{app:poisson_distrib}. In the following, we mostly consider the enclosing volumes $V$ to be $d$-dimensional balls. 
Let us denote by $N(r)$ the average number of particles in a $d$-dimensional ball of radius~$r$. 
We have the expression 
\begin{equation}
N(r)=\frac{\pi^{d/2}}{\Gamma(\tfrac d2+1)}\rho r^d = {\left(\frac{r}{\lambda_d}\right)}^d,
\label{eq:Nr}
\end{equation}
with $\lambda_d=\big(\pi^{d/2}\,\rho/\Gamma(1+\tfrac d2)\big)^{-1/d}$,
therefore the Poisson distribution of the number~$n$ of particles in a ball of radius~$r$, when particles are point-like, is 
\[ P_n(r) = \frac{N(r)^n}{n!}\e^{-N(r)}.\]
We write $w_1(r)$ to be the probability density of finding the nearest neighbor of a particle at a distance~$r$. We then have 
\[ \int_0^r w_1(r')\d r' = 1 - P_0(r).\]
Differentiating this expression, we find \cite{chandra43}
\begin{equation}
    w_1(r)=-\der{P_0}{r}=N'(r)\e^{-N(r)}.
\end{equation}
In Appendix \ref{app:finite_size}, we show similar derivation but for particles with finite size. Similarly, the distribution of the distance of the $n^{\text{th}}$ nearest neighbor, $w_n(r)$, satisfies the equation
\[ \int_0^rw_n(r')\d r'=\sum_{k=n}^\infty P_k(r)\]
such that we have after differentiating with respect to $r$
\[ w_n(r) = \sum_{k=n}^\infty \der{P_n}{r} = 
  \sum_{k=n}^\infty \left[N'(r) \frac{N(r)^{k-1}}{(k-1)!}\e^{-N(r)}-\frac{N(r)^k}{k!}N'(r) \e^{-N(r)} \right].\]
The series is telescoping, so that there remains the result
\begin{equation}
    w_n(r)=N'(r)\frac{N(r)^{n-1}}{(n-1)!}\e^{-N(r)}.
    \label{eq:wn}
\end{equation}
In Appendix \ref{app:properties_wn}, we compute key properties of the $w_n(r)$ distribution, including its mean and variance. An alternative derivation of this distribution, based on the spatial properties of a set of particles within a spherical $d$-dimensional volume, is presented in Appendix \ref{app:alternative_nth2}.
\subsection{Joint distributions}
We can straightforwardly extend the preceding result to the joint distribution of the distances to the first, second, and up to the $n$-th nearest neighbors, $w^{(n)}(r_1,,r_2,\dots, r_n)$.
To begin, consider the joint distribution of the two nearest neighbors, $w^{(2)}(r_1,,r_2)$, with $r_1 < r_2$.
The probability density for $r_2$ is given by $w_2(r_2)$, and for a fixed value of $r_2$, the nearest neighbor is uniformly distributed within the sphere of radius $r_2$, in accordance with the Poisson law, with density $N'(r_1)/N(r_2)$.
This density is properly normalized, since the integral of $N'(r_1)$ from 0 to $r_2$ equals $N(r_2)$. This leads to the final result:
\[ w^{(2)}(r_1, \, r_2)=N(r_2)N'(r_2)\e^{-N(r_2)}\frac{N'(r_1)}{N(r_2)} = N'(r_1)N'(r_2)\e^{-N(r_2)}. \]

This result can be generalized by considering $r_n$, the largest of the $n$ distances, and its distribution $w_n(r_n)$.
For a fixed value of $r_n$, the remaining $n-1$ distances are uniformly distributed within the sphere of radius $r_n$.
If these $n-1$ distances are treated as independent, each is distributed between $0$ and $r_n$ according to the density $N'(r_i)/N(r_n)$.
In our definition, these distances are ordered ($r_1 \leq r_2 \leq \cdots \leq r_{n-1}$), which requires including a normalization factor of $(n-1)!$.
We get that
\[
w^{(n)}(r_1,\,r_2,\,\dots\,r_n)=N'(r_n)\frac{N(r_n)^{n-1}}{(n-1)!} \e^{-N(r_n)} \,
 \frac{N'(r_1)}{N(r_n)} \frac{N'(r_2)}{N(r_n)} \dots \frac{N'(r_{n-1})}{N(r_n)}\,(n-1)!,
 \]
 and after some simplifications
 \begin{align}
    w^{(n)}(r_1,\,r_2,\,\dots\,r_n)&=N'(r_1)N'(r_2)\dots N'(r_n)\e^{-N(r_n)}
    \qquad (r_1\leq r_2\leq\cdots \leq r_n),\\
   &= \mbox{e}^{-N(r_n)}\prod_{i = 1}^{n}N'(r_i) \qquad (r_1\leq r_2\leq\cdots \leq r_n).
    \label{eq:wN}
\end{align}
It is worth noting that the expected relation between $w^{(n+1)}$ and $w^{(n)}$ is satisfied:
\[
\int_{r_n}^\infty w^{(n+1)}(r_1,\,r_2,\,\dots, r_{n+1}) \, \mathrm{d}r_{n+1} = w^{(n)}(r_1,\,\dots, r_n).
\]
The joint distribution of a 2-particle subset of these distances can be obtained by partial integration, such as
\[ w_{m,n}(r_m,\,r_n)=\int\dots\int \Big(\prod_{\substack{1\leq\, j\,< n\\j\neq m}}\d r_j\Big) w^{(n)}(r_1,\,\dots\, r_{n-1})\]
The integral is elementary and yields 
\[ w_{m,n}(r_m,\,r_n) = \frac{N(r_{m})^{m-1}}{(m-1)!} N'(r_m) 
  \frac{\big(N(r_n)-N(r_m)\big)^{n-m-1}}{(n-m-1)!} N'(r_n)\e^{-N(r_n)}. \]
It is interesting to observe that the probability density of $r_{n+1}$ conditioned by $r_n$ is independent of $n$. 
We have indeed 
\[ w_{n+1|n}(r_{n+1}\,|\,r_n) = \frac{w_{n,n+1}(r_n,\,r_{n+1})}{w_n(r_n)} 
   = N'(r_{n+1}) \e^{N(r_n)-N(r_{n+1})} \; \Theta(r_{n+1}-r_n). \]
This result represents the probability density of the distance $r'$ to the next nearest neighbor beyond $r$, denoted $w^{\text{next}}_r(r')$.  
One observes that the distribution $w_r^{\text{next}}$ becomes increasingly peaked as $r$ increases.

The correlation coefficient $\rho(n)$ between the two random variables $r_n$ and $r_{n+1}$ is not strongly affected by the spatial dimension $d$.  
Moreover, the radial coordinates $r_n$ and $r_{n+1}$ become increasingly correlated as the index $n$ grows.  
This reflects the accumulation of particles within layers of arbitrary thickness at distances $R_n \propto n^{1/d} \lambda_d$ surrounding the $d$-sphere centered on the test particle, which arises purely from geometrical effects.

\section{Gravitational forces in homogeneous random media}
\label{sec:gravity_homogeneous_media}
Up to this point, we have used order statistics to characterize the spatial properties of neighboring particles that are randomly distributed within an infinite volume of constant density, relative to a given test particle. Building on these results for the statistical properties of nearest-neighbor distances, we next examine how they inform the study of the total Newtonian gravitational force acting on a test particle in a random homogeneous medium—described by the Holtsmark distribution in dimension $d=3$, presented in Section \ref{sec:Holtzmark_distrib}. In Section \ref{sec:n_nearest_neighbor_pdf_force}, we analyze the contribution of the nearest neighbors to the total gravitational force, and in Section \ref{sec:non-finite_variance}, we investigate the non-finite variance of the Holtsmark distribution using order statistics.
\subsection{The Holtsmark distribution}
\label{sec:Holtzmark_distrib}
In $d = 3$, we denote the volume element by $v_3 = V$.  
We consider a test particle of mass $m$ located at $\mathbf{x}$ in a gas of identical particles, each of mass $m$, labeled by $i$.  
The total gravitational force $\mathbf{F}$ exerted on the test particle by its neighbors can be obtained by summing the contributions from all particles at their respective positions $\mathbf{x}_i$, and is given by:
\begin{equation}
    \textbf{F}(\textbf{x}) = \sum\limits_{i = 1}^N\frac{Gm^2}{|\textbf{x}_i - \textbf{x}|^3}(\textbf{x}_i - \textbf{x}) = \sum\limits_{i = 1}^N \textbf{F}_i,
\end{equation}
with the gravitational constant $G = 6.674 \times 10^{-11}\ \mbox{m}^3\,\mbox{kg}^{-1}\,\mbox{s}^{-2}$.  
We can provide a general expression for the probability density function $W(\mathbf{F})$, which gives the probability of finding the gravitational force on the test particle within a small volume $\mbox{d}^3\mathbf{F}$ around $\mathbf{F}$, due to all $N$ particles inside a sphere centered on it.  
This probability can be expressed as \cite{Pietronero_2002}:
\begin{equation}
W(\textbf{F}) = 
\int_{V}...\int_{V}\mbox{d}\textbf{x}^{3N}f(\textbf{x}_1,...,\textbf{x}_N)\delta^D\left(\textbf{F} - \sum\limits_i^N \textbf{F}_i\right),
\end{equation}
where $f(\mathbf{x}_1, \dots, \mathbf{x}_N)$ is the joint probability density function for finding the $N$ particles at positions $\mathbf{x}_1, \dots, \mathbf{x}_N$. For a Poisson-like distribution of particles, the joint probability function $f$ factorizes into the product of constant probability densities for each particle, given by $1/V$, such that:
\begin{equation}f(\textbf{x}_1,...,\textbf{x}_N)\mbox{d}\textbf{x}^{3N} = \frac{\mbox{d}\textbf{x}^{3N}}{V^N}.
    \label{jointdensity}
\end{equation}
The statistical distribution of the modulus $F = |\textbf{F}|$ is known as the Holtsmark distribution \cite{holtsmark19,chandra43,Sylos_Labini_2008,chavanis09} and is given by (see the calculation details in Appendix \ref{app:holsmark})
\begin{equation}
W_H(F) = \frac{2}{\pi F}\int_0^\infty\mbox{d}x(x\sin{x})\exp[ - (x/\beta)^{3/2}].
\label{eq:holtsmark}
\end{equation}
where $\beta = F/F_0$ and
\begin{equation}
F_0 = 2\pi\left(\frac{4}{15}\right)^{2/3}\frac{Gm^2}{\lambda^2},
\label{F0}
\end{equation} 
with $\lambda = \rho_0^{-1/3}$. The distribution is represented in dashed line in Figure \ref{fig:w_distrib}.
The average force is given by the integral
\begin{equation}
\langle F\rangle = F_0^2\int_0^\infty\beta W_H(\beta)\mbox{d}\beta = \frac{4 F_0}{\pi}\Gamma\left(\frac{1}{3}\right) \approx 3.412F_0.
\end{equation}
where $\Gamma$ is the Gamma function.  
A key property of this distribution is that its variance is formally infinite.  
This occurs because the asymptotic behavior of $W_H(F)$ for large $F$ follows a power law, specifically $W_H(F) \sim F^{-5/2}$. While this decay is sufficient to ensure a well-defined mean force, the second moment $\langle F^2 \rangle$ diverges, implying that the variance of the Holtsmark distribution diverges.

\subsection{Nearest neighbor contribution to the total gravitational force}
\label{sec:n_nearest_neighbor_pdf_force}
In the previous section, we analyzed the statistical properties of the total gravitational force acting on a test particle due to an infinite number of neighboring particles randomly distributed with uniform density. Building on the results obtained in Section \ref{sec:stat_distances}, we now examine the contribution of each nearest neighbor to the total gravitational force.

\begin{figure}

\centering\includegraphics[width=0.8\linewidth]{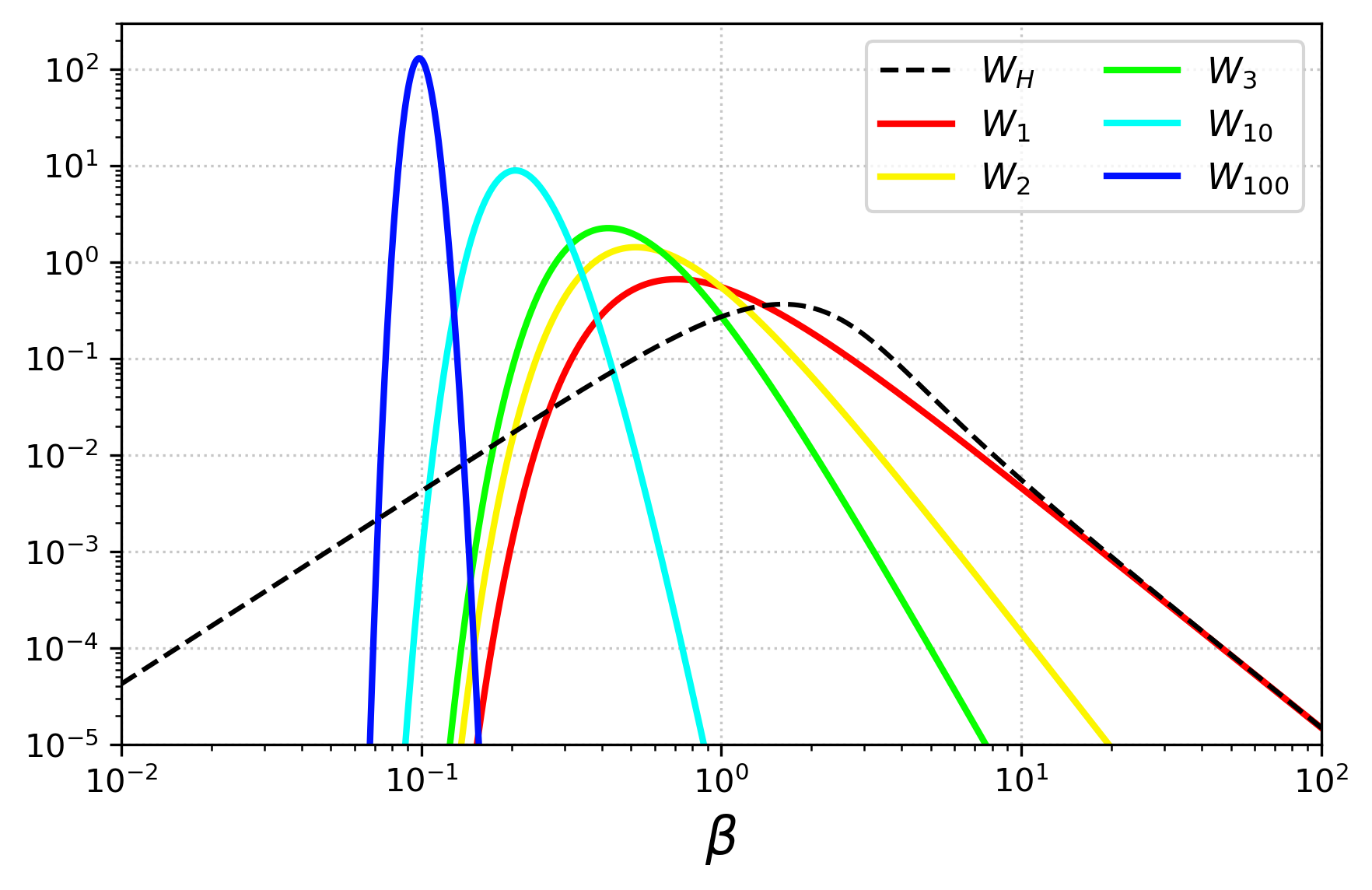}
    \caption{The Holtsmark distribution in Eq. \eqref{eq:holtsmark} is represented in dashed line. The $n$-th nearest neighbor gravitational force probability distributions in Eq. \eqref{eq:Wn} for $n=\{1,2,3,10,100\}$ are represented in colored full-lines.}
    \label{fig:w_distrib}
\end{figure}

The probability density $W_n(F)$ of the force modulus $F_n = |\mathbf{F}_n|$ for the $n$-th nearest neighbor can be directly derived from its spatial distribution $w_n(r)$ through the relation $W_n(F)\, \mathrm{d}F = w_n(r)\, \mathrm{d}r$.  
The resulting $W_n(F)$ is given by the expression:
\begin{equation}
    W_n(\beta) = \frac{[5/2\sqrt{2\pi}]^n}{F_0 \, \Gamma(n)}\frac{3}{2}\beta^{-(1 + 3n/2)}\mbox{exp}\left(-\frac{5\beta^{-3/2}}{2\sqrt{2\pi}}\right),
    \label{eq:Wn}
\end{equation}
where we consider $\beta = F/F_0$ and $F_0$ to be the typical force defined in the case of the Holtsmark distribution in Eq.~\eqref{F0}. The $W_n$ distributions for $n=\{1,2,3,10,100\}$ are shown in colored lines in Fig. \ref{fig:w_distrib}.  
The average force modulus $\langle F \rangle_{n}$ is given by
\begin{equation}
    \langle F_n\rangle = \frac{F_0}{\Gamma(n)}\left(\frac{5}{2\sqrt{2\pi}}\right)^{2/3}\Gamma\left(n - \frac{2}{3}\right).
    \label{eq:meanFn}
\end{equation}
As expected in dimension $d=3$, the average force modulus $\langle F\rangle_n$ decreases to 0 when $n$ increases. The moment of order $2$ reaches the following expression for $n > 1$ (not finite for $n = 1$)
\begin{equation}
   \langle F^2_n\rangle = \frac{F_0^2}{\Gamma(n)}\left(\frac{5}{2\sqrt{2\pi}}\right)^{4/3}\Gamma\left(n - \frac{4}{3}\right).
\end{equation}
More generally, for $k > 2$, the $k$-th moment is proportional to $\langle F^k_n\rangle \propto \Gamma(n - 2k/3)/\Gamma(n)$, which is finite and positive for $n > 2k/3$.

The corresponding dispersion $\sigma_{W,n}$ is given by:
\begin{equation}
    \sigma_{W,n}^2 = F_0^2\left(\frac{5}{2\sqrt{2\pi}}\right)^{4/3}
     \left[\frac{\Gamma(n-4/3)}{\Gamma(n)} - \frac{\Gamma(n-2/3)^2}{\Gamma(n)^2}\right].
    \label{eq:sigma_Wn}
\end{equation}
Using the expansion, as $n\to\infty$, 
\[ \frac{\Gamma(n+x)}{\Gamma(n)} = n^{x} \left( 1 + \frac{x(x-1)}{2n} + O\!\left(\frac{1}{n^{2}}\right) \right), \]
we obtain the large~$n$ equivalent $\sigma_{W,n}\approx \left(\frac{5}{3\sqrt{3\pi}}\right)^{2/3}F_0\,n^{-7/6}$.

We can investigate the case $n = 1$ where the distribution of force takes the form
\begin{equation}
   W_1(F) = \frac{15\beta^{-5/2}}{4\sqrt{2\pi}F_0}\exp\left(-\frac{5\beta^{-3/2}}{2\sqrt{2\pi}}\right),
\end{equation}
with the average force modulus $\langle F_1\rangle$
\begin{equation}
\langle F_1\rangle = \left(\frac{5}{2\sqrt{2\pi}}\right)^{2/3}\Gamma\left(\frac{1}{3}\right)F_0 \approx 2.674F_0.
\end{equation}
We see that, in Fig. \ref{fig:w_distrib}, the tail of the $W_1$ distribution in the $\beta \gg 1$ regime coincides with the shape of the Holtsmark distribution. For $F \rightarrow \infty$, reporting the first two terms of the Taylor expansion of $W_1(F)$
\begin{equation}
W_1(F) \sim \frac{15}{8F_0}\left(\frac{2}{\pi}\right)^{1/2}\beta^{-5/2} - \frac{75}{16\pi F_0}\beta^{-4}.
\end{equation}
We get exactly the first term of the Holtsmark distribution Taylor expansion for $\beta \gg 1$, given by
\begin{equation}
W_H(F) \approx \frac{15}{8F_0}\left(\frac{2}{\pi}\right)^{1/2}\beta^{-5/2} - \frac{24}{\pi F_0}\beta^{-4}.
\label{Hlarge}
\end{equation}
This implies that in the regime $F \rightarrow \infty$, the largest contribution to the force on the test particle comes from the nearest neighbor. For low $F$, the exponential term cannot be expanded as a Taylor series.\\

More generally, for any $n$-th nearest neighbor, we get the  Taylor expansion for $W_n(F)$ for $\beta \gg 1$
\begin{align}
    W_n(\beta)&= \frac{[5/2\sqrt{2\pi}]^n}{F_0(n-1)!}\frac{3}{2}\sum\limits_{k = 0}\frac{(-1)^k}{k!}\left(\frac{5}{2\sqrt{2\pi}}\right)^{k}\beta^{-(1 + 3(n+k)/2)}\\
    &\approx  \frac{[5/2\sqrt{2\pi}]^n}{F_0(n-1)!}\frac{3}{2}\left[\beta^{-(1+3n/2)} - \frac{5}{2\sqrt{2\pi}}\beta^{-(5/2 + 3n/2)}\right].
\end{align}
At first order, we retrieve for a given index $k$ the exponent of $\beta$ in the series development of the Holtsmark distribution for large $\beta \gg 1$, given by the general term $\beta^{-b_n}$ with $b_n = 1 + 3n/2$ in
\begin{align}
    W_H(F) &= \frac{2}{\pi F_0}\sum\limits_{n = 0}\frac{(-1)^{n + 1}}{n!}\sin\left(\frac{3\pi n}{4}\right)\Gamma\left(\frac{3n}{2} + 2\right)\beta^{-(1+3n/2)}.
\label{Hlargen}
\end{align}
\subsection{Non-finite variance of the Holtsmark distribution}
\label{sec:non-finite_variance}
We re-introduce the random variable $\mathbf{F}$, defined as the total gravitational force exerted on a test particle from a set of $N$ neighboring particles (taken to be the $N$ nearest neighbors), such as
\begin{equation}
    \mathbf{F} = \sum\limits_{i = 1}^N \mathbf{F}_i\
\end{equation}
The squared total force, $|\mathbf{F}|^2$ is given by
\begin{equation}
    |\mathbf{F}|^2 = \sum\limits_{i,j = 1}^N \mathbf{F}_i\cdot\mathbf{F}_j = F_0^2\sum\limits_{i,j = 1}^N \beta_i\beta_j\, \times\,\textbf{u}_i \cdot \textbf{u}_j,
\end{equation}
Such as the second moment of its squared modulus is given by
\begin{equation}
    \langle|\mathbf{F}^2|\rangle =  F_0^2\sum\limits_{i,j = 1}^N \langle \beta_i\cdot\beta_j \rangle \,\times \langle \textbf{u}_i \cdot \textbf{u}_j\rangle,
\end{equation}
where we assume that $\textbf{u}_i$ and $\beta_i$ are independent random variables. By definition, when $N\rightarrow+\infty$, $\langle|\mathbf{F}^2|\rangle$ corresponds to the second moment of the distribution $W_H(F)$. Taking $\langle \textbf{u}_i \cdot \textbf{u}_j\rangle$ to be $2\delta_{ij}^K/6$ in dimension $d=3$ (for random distribution of the nearest neighbors) we get that
\begin{equation}
    \langle|\mathbf{F}^2|\rangle\propto  \frac{2}{6}\sum\limits_{i = 1}^N \langle\beta_i^2\rangle
     = \frac{2}{6}\left[\sum\limits_{i = 1}^N \langle\beta_i\rangle^2+ \left(\frac{\sigma_{W,i}}{F_0}\right)^2 \right].
     \label{eq:mean_beta2}
\end{equation}
where $\sigma_{W,i}$ (given in Eq. \eqref{eq:sigma_Wn}) is the variance of the force exerted by the $i$-th nearest neighbor, and $\langle\beta_i\rangle$ (given in Eq. \eqref{eq:meanFn}) corresponds to its average value. The first term in Eq.~\eqref{eq:mean_beta2} gives  
\begin{equation}
    \sum\limits_{i = 1}^N \langle \beta_i \rangle^2 \propto \sum\limits_{n = 1}^N n^{-4/3},
\end{equation}
which converges as $N \rightarrow \infty$ (the above sum converges to the Riemann zeta function $\zeta(s)$ evaluated at $s=4/3$). For the second part of the sum in Eq.~\eqref{eq:mean_beta2}, we have
\begin{equation}
     \sum\limits_{i = 1}^N \sigma_{W,i}^2 = \sigma_{W,1}^2+ \sum\limits_{i = 2}^N \sigma_{W,i}^2.
     \label{eq:sum_sigmaW}
\end{equation}
We have that $\sigma_{W,1}^2$ in Eq. \eqref{eq:sum_sigmaW} is not finite, since $\langle F^2\rangle_1$ diverges as demonstrated in Section \ref{sec:n_nearest_neighbor_pdf_force}. For the second term in Eq. \eqref{eq:sum_sigmaW}, we get that
\begin{equation}
    \sum\limits_{i = 2}^N \sigma_{W,i}^2\propto \sum\limits_{n = 2}^N n^{-7/3},
\end{equation}
which converges in the limit $N \rightarrow \infty$.  
From the above, we found that $\langle|\mathbf{F}^2|\rangle$ diverges because of the first nearest neighbor contribution ($n = 1$) to the total variance, which has $\sigma_{W,1}$ infinite.  
Hence, the non-finite dispersion of the distribution $W(|\beta|)$ arises entirely from the contribution of the nearest neighbor to the variance $\sigma_{W,1}$, as noted in \cite{chavanis09} for $d = 3$ (we remark that the variance of the Holtsmark distribution is finite in $d = 1$ and diverges logarithmically with $N$ in $d = 2$.).

We can interpret that knowing that for $n > 1$, the pair correlation for the radial coordinate, $\rho(n)$, approaches 1.  The $n$-th nearest neighbors then accumulate at approximately the same distance, reducing their contribution to the total force modulus.  
In contrast, the first nearest neighbor is much closer and, on average, exerts a larger force on the test particle. The fluctuations of the force associated with the distribution $W_1(F)$ are also greater than those of any other neighbors and dominate statistically, with weights much larger than the distributions $W_{n > 1}(\beta)$ when $\beta \gg 1$.


\section{Conclusions}
\label{sec:conclusion}
The Holtsmark distribution provides insight into the collective statistics of the gravitational force in a random homogeneous gas with a Poisson-like spatial distribution of point particles.  
We have discussed the statistical properties of particle locations in a $d$-dimensional space of constant density $\rho_0$, considering the neighborhood of an arbitrary center particle.  
In particular, we generalized the joint probability density function for finding the $N$ nearest neighbors at their respective positions.  
The analysis of the joint probability density of particle locations shows that the radial coordinates of two nearest neighbors with successive indices $n$ become strongly correlated as $n$ increases.  

In three dimensions, we analyzed the contribution of the $n$-th nearest neighbor to the total gravitational force exerted on a test particle.  
We observed that the formally divergent variance of the Holtsmark distribution arises entirely from the force contribution of the first nearest neighbor.  

Future studies could extend this approach by using the joint probability distribution for any number of neighbors to investigate instabilities in a gas of gravitationally bound particles with an initially homogeneous spatial distribution, which can lead to the formation of coherent structures such as filaments.  
Additionally, the effect of a continuous mass distribution within the gas could be incorporated to study more general statistical properties of gravitational fields.

\appendix
\section{Elementary properties of a Poisson distribution}
\label{app:poisson_distrib}
Consider a spatial distribution of points in a $d$-dimensional space, characterized with density~$\rho$
and consider a small $d$-volume $\d v$, such that $\rho\,\d v\ll1$. The number of particles present in this volume is
most likely equal to zero. Denoting by $p_n(\d v)$ the probability of having~$n$ particles in a volume~$\d v$, we have
\[ 1-p_0(\d v)\ll1.\]
The average number of particles in $\d v$, denoted by $\avg{n(\d v)}$, is by definition equal
to $\rho\,\d  v$, and it is also expressed as 
\[ \avg{n(\d v)} = p_1(\d v)+2 p_2(\d v) +\cdots  \]
Since the positions of particles are uncorrelated, we have $p_2(\d v) \approx p_1(\d v)^2 \ll p_1(\d v)$.
In other words, $p_2(\d v)$ is of second order in $\rho\,\d v$, and so are $p_n(\d v)$ for $n\geq 2$.
We have therefore, at first order in $\rho\,\d v$, the equality $p_1(\d v) = \rho\,\d v + o(\rho\,\d v)$. 
Consider now a volume~$v$ and a slightly larger volume~$v+\d v$, with $\d v\ll v$ (and $\rho\,\d v\ll1)$.
The relation 
\[ p_0(v+\d v)=p_0(v)\,p_0(\d v) = p_0(v)\, \big(1-\rho \d v\big)\]
expresses the fact that there are no particles in the volume $v+\d v$ 
if and only if there are no particles in~$v$ and no particles
in $\d v$ either. From this equality, we obtain the differential equation for~$p_0$
\[ \der{p_0}{v} = - \rho \, p_0.\]
Since $p_0(0)=1$, we have the solution 
\begin{equation}
    p_0(v)=\e^{-\rho v}.
\end{equation}
The Poisson statistics are obtained using a recurrence with $p_n(v)=g_n(v)\e^{-\rho v}$, where $g_n$ is a polynomial.
We have as before 
\[ p_{n+1}(v+\d v)=p_{n+1}(v)p_0(\d v)+p_n(v)p_1(\d v) = p_{n+1}(v)\big(1-\rho \d v\big)+p_n(v)\rho \d v.\]
This yields the differential equation
\[ \der{p_{n+1}}{v} = - \rho \, p_{n+1} + \rho \, p_n\]
which reduces to 
\[ \der{g_{n+1}}{v} = g_n(v).\]
Since $g_0=1$, we immediately obtain $g_n(v)=(\rho v)^n/n!$ and the classical Poisson distribution of parameter~$\rho\,v$.
\begin{equation}
    p_n(v)=\frac{(\rho v)^n}{n!}\e^{-\rho v}.
    \label{eq:Poisson}
\end{equation}
Note that we have indeed 
\[ \avg{n(v)}=\sum_{k=0}^\infty k\, p_k(v)=\rho v.\]

\section{Nearest-neighbor distribution of particles with finite sizes}
\label{app:finite_size}
Let us now turn to the case of particles with finite size. The presence of excluded volume effects alters the distribution of distances to the nearest neighbors, and this dependence is generally influenced by the particle shape. In what follows, we restrict our analysis to spherical particles of radius $a$. We denote by $\rho$ the spatial density of particles, defined as the average number of particles per unit volume. Throughout this discussion, the number of particles within a volume refers to the number of particle centers contained in that volume. We adopt the notation
\begin{equation}
    \alpha_d=\frac{\pi^{d/2}}{\Gamma\big(\tfrac{d}{2}+1\big)}
\end{equation}
such that the volume of the $d$-dimensional ball of radius~$r$ is $\alpha_d\,r^d$. Let us start with the study of $p_n(r)$, the probability that a sphere of radius~$r$ contains exactly $n$~particles.
In $d$ dimensions, the packing of spheres has a limit fraction~$\kappa_d$ (for instance $\kappa_2=\frac{\pi}{2\sqrt{3}}$ and 
$\kappa_3=\tfrac{\pi}{3\sqrt{2}}$, see \citep{Hales_2005}). The density~$\rho$ is therefore bounded 
\begin{equation}
    \rho \leq \frac{\kappa_d}{\alpha_d} a^{-d}. 
\end{equation}
Since no excluded volume effect is involved when considering less than two particles, we can infer that
for $r<2a$
\begin{equation}
  p_1(r) = \Big(\frac{r}{\lambda_d}\Big)^d \qquad(r<2a)
  \label{fs:p1}
\end{equation}
Excluded volume effects imply that for $n\geq2$ and $r<a$, $p_n(r)=0$.
Since we have the equality 
\[ p_0(r) + p_1(r) + \cdots = 1, \]
we conclude that for $r<2a$, we have
\begin{equation}
p_0(r)= 1-p_1(r) = 1-\Big(\frac{r}{\lambda_d}\Big)^d \qquad(r<2a).
\label{fs:p0}
\end{equation}
We observe that the finite size slightly modifies the probability~$p_0$. As long as $a\ll\lambda_d$
(low density limit), this modification is small. However, for $a\lesssim\frac12\lambda_d^{\phantom{d}}\,\kappa_d^{1/d}$,
$p_0(r)$ is significantly larger than $\e^{-(r/\lambda_d)^d}$.

Let us now compute $p_0(r)$ for $r\geq a$. We have the relation
\[ p_0(r+\d r) = p_0(r) \, \Phi_d(r)\d r \]
where $\Phi_d(r,\, \d r)$ is the probability that there are no particles the shell
\[ \Sigma_d(r,\,\d r)= \big\{\vr \in \mathbf{R}^d,\, r\leq \|\vr\|\leq r+\d r\big\}. \]
The probability that $\Sigma_d(R,\,d r)$ contains at least one particle is $d\,\alpha_d \rho r^{d-1}\d r$,
and the probability that it contains two particles is of the second order in~$\d r$, so we can conclude
that at the first order in $\d r$, $\Phi_d(r,\,d r)=d\,\big(\frac{r}{\lambda_d}\big)^{d-1}\frac{\d r}{\lambda_d}$.
It follows that $p_0$ is determined by the differential equation
\[ \der{p_0}{r}= - d\,\alpha_d \frac{r^{d-1}}{\lambda_d^d} \, p_0(r)\qquad(r\geq 2a)\]
with the boundary condition $p_0(a)=1-\big(\frac{a}{\lambda_d}\big)^d$. 
The solution is
\begin{equation}
    p_0(r) = \left(1-\big(\frac{2a}{\lambda_d}\big)^d\right) \exp\left[\frac{2^da^d-r^d}{\lambda_d^d}\right].
\end{equation}

\section{The $n$\textsuperscript{th} nearest neighbor distance distribution}
\subsection{Properties of the distribution}
\label{app:properties_wn}
Let us consider the distributions $w_n$ of the distance between the $n^{\text{th}}$ nearest neighbor and the reference particle.
The maximum of the distribution is located at the unique zero of~$\der{w_n}{r}$. For $n\geq2$
\[ \der{w_n}{r} = \frac{N^{n-2}}{(n-2)!}\e^{-N}\left(N''\frac{N}{n-1}+N'^2-N'^2\frac{N}{n-1}\right)\]
yielding the distance $r_n^{\max}$ such that 
\[ N(r_n^{\max})=n-\frac{1}{d}\quad\text{or}\quad r_n^{\max}=\lambda_d \big(n-\tfrac{1}{d}\big)^{1/d}.\]
and the average of this distance is 
\[ \avg{r_n}=\int_0^\infty N'(r) \e^{-N(r)} \frac{N^{n-1}}{(n-1)!} \, r\, \d r
            =\int_0^\infty \e^{-N} \frac{N^{n-1}}{\Gamma(n)} \, \lambda_d \, N^{1/d} \;\d N
            =\lambda_d\,\frac{\Gamma\big(n+\tfrac{1}{d}\big)}{\Gamma(n)}.\]
Similarly, we have the $k^{\text{th}}$ order  
\[ \avg{r_n^k} = \lambda_d^k \, \frac{\Gamma\big(n+\tfrac{k}{d}\big)}{\Gamma(n)}
  \underset{n\to\infty}\sim \lambda_d^k \; n^{k/d}. \]
In particular $\sigma_n^2=\avg{r_n^2}-\avg{r_n}^2\underset{n\to\infty}{\sim} 
  \left(\dfrac{\lambda_d}{d}\right)^d\, n^{-1+2/d}.$
  \subsection{Alternative derivation: Distribution of the $n$-th nearest neighbor from a set of particles in a spherical $d$-volume}
\label{app:alternative_nth2}
We can obtain the same result as in Eq.~\eqref{eq:wn} by considering a finite spherical $d$-dimensional volume of radius $R$, given by $v_d(R) = \alpha_d R^d / d$. 
Let $r$ denote the radial coordinate of a particle within the sphere, and define the reduced variable $s = r/R \in [0,1]$. For a free particle in the system, the probability $p(s)\, ds$ of finding it at $s$ is given by the ratio of the volume of the spherical shell between $r$ and $r + dr$ to the total volume $v_d$, which can be written as

$$
    p(s) = \left|\frac{\mbox{d}r}{\mbox{d}s}\right|\frac{\alpha_d r(s)^{d-1}}{v_d} = ds^{d-1} 
$$
The volume is now filled with $N$ particles with average density $\rho_0 = N/v_d$. For each realization of $N$ particles, we order the positions $\{s_i\}_{1\leq i\leq N}$ in a list. Let's note that the formalism given by the variable $s = r/R$ generalizes the equivalent radius of the system to 1 and makes statistical properties depend only on the total number of particles $N$ within the volume $v_d$. 

It is possible to deduce an evaluation of $w_n$ by considering order statistics results. We consider the new random variable $\mathcal{X}$, which corresponds to the $n$-th value of a set of $N$ elements $\{s_i\}_{1\leq i\leq N}$ distributed following $p(s)$ and ordered in a list. We define the probability density function $f_n$ that verifies the equality $\mbox{P}_n(\mathcal{X} \in [x,x+\mbox{d}x[) =  f_n(x)\mbox{d}x$. We can write the general expression for $f_n(x)$
$$
f_n(x) 
     \sim p(x)\left[\int_0^xp(s)\mbox{d}s\right]^{n-1}\left[\int_x^1p(s)\mbox{d}s\right]^{N-n}\sim x^{dn-1}(1-x^d)^{N-n}.
$$
Now we investigate the change of variable $x \rightarrow r$ to get the probability density function $w_n(r)$ that must verify $f_n(x)\mbox{d}x = w_n(r)\mbox{d}r$. We can express a general form for $w_n$ such as:
$$
    w_n(r) \sim \left(r/R\right)^{dn-1}[1 - (r/R)^d]^{N-n} \rightarrow \left(r/R\right)^{dn-1}\exp[-(N-n)(r/R)^d]
$$
Where the second part of the above equation is obtained by taking the limit of infinite system $N, R \rightarrow \infty$ and $N/v_d = \rho_0$, by developing $ 1 - x \approx \exp[-x]$ when $x = r/R \ll 1$. For $n$ small enough compared to $N$, the difference $N-n \approx N \gg 1$, then we get the exponent $(N-n)(r/R)^{d} \approx N(r)$ where $N(r)$ is given in Eq. \eqref{eq:Nr}. We compute the appropriate normalization from the integral of $w_n$ from 0 to $\infty$. We get
$$
    w_n(r) = \frac{1}{(n-1)!}N(r)^{n-1}N'(r)\mbox{e}^{-N(r)}.
$$
We retrieve the same result as in Eq. \eqref{eq:wn} by considering the statistical properties of ordered list elements distributed following the density $p(s)$.

\section{Holtsmark distribution: calculation details}
\label{app:holsmark}
The distribution of force is then given by the integral over $N$ volumes
\begin{equation}
W_H(\textbf{F}) = \int_{V}...\int_{V}\left[\frac{\mbox{d}\textbf{x}^{3N}}{V^N}\right]\delta\left(\textbf{F} - \sum\limits_i^N \textbf{F}_i\right).
\end{equation}

By giving the expression of $W_H$ as an inverse Fourier transform, we get
\begin{equation}
W_H(\textbf{F}) = \frac{1}{(2\pi)^3}\int_{\Omega(\textbf{k})} \mbox{d}^3\textbf{k} A(\textbf{k})\exp(-i\textbf{k}\cdot\textbf{F}),
\end{equation}
where $A(\textbf{k})$ is given by
$$
A(\textbf{k}) = \left[1 - \frac{\rho_0}{\rho_0 V} \int_V\mbox{d}^3\textbf{x}[1-\exp(i\textbf{k}\cdot\textbf{F}_i(\textbf{x}))]\right]^{\rho_0 V}.
$$
By using the definition of the exponential function, 
for infinite volume $v_d$, infinite number of particle $N$, and constant density $\rho_0$, we get
$$
    A(\textbf{k}) = \lim_{N\to\infty}\left[1 - \frac{\rho_0 C(\textbf{k})}{N} \right]^N = \exp(-\rho_0 C(\textbf{k})),
$$
with
$$
C(\textbf{k}) = \int_V \mbox{d}^3\textbf{x}[1 - \exp(i\textbf{k}\cdot\textbf{F}_i(\textbf{x}))].
 \label{CkH}
$$
By considering an arbitrary axis given by the vector \textbf{k}, integrating over all the directions gives
$$
    C(k) = 2\pi (kGm^2)^{3/2}\int_0^\infty\frac{\mbox{d}u}{u^{5/2}}\left[u - \sin(u)\right] = \frac{4}{15}(2\pi k G m^2)^{3/2}.
$$
where we used 
$$
\int_0^{\infty}x^\alpha\sin x \mbox{d}x = -\cos\left(\frac{\alpha \pi}{2}\right)\Gamma(1 + \alpha).
\label{intsin}
$$
to identify 
$$
    \int_0^\infty\frac{\mbox{d}u}{u^{5/2}}\left[u - \sin(u)\right] = \int_0^\infty\frac{\sin(u)}{u^{3/2}}\mbox{d}u = \frac{4\sqrt{2\pi}}{15}.
$$
The Holtsmark distribution is given by the integral 
$$
W_H(\textbf{F}) = \frac{1}{(2\pi)^3}\int_{\Omega(\textbf{k})}\mbox{d}^3\textbf{k}\exp[-i\textbf{k}\cdot\textbf{F} - \frac{4\rho_0}{15}(2\pi k G m^2)^{3/2}].
$$
By considering, as before, an arbitrary axis oriented along $\textbf{F}$ giving $\textbf{k}\cdot\textbf{F} = kF\cos{\theta}$, integrating over the directions and radial variable $k$ gives the expression 
\begin{equation}
W_H(\textbf{F}) = \frac{1}{2\pi^2F^3}\int_0^\infty \mbox{d}x (x\sin{x})\exp\left[-(x/\beta)^{3/2}\right]
\label{eq:Holtsmark_distrib_angl}
\end{equation}

\end{document}